\documentclass[letterpaper,nofootinbib,twocolumn,floatfix,showpacs,prd]{revtex4}
\usepackage{graphics,rotating,multirow,bm,amsmath, amssymb}

\def\etal{{\frenchspacing\it et al.}}

\def\be{\begin{equation}}
\def\ee{\end{equation}}
\newcommand{\mr}[1]{ \mathrm{ #1}}

\begin{document}

\title{Optimized supernova constraints on dark energy evolution}

\author{Christian Stephan-Otto\footnote{Electronic address: \texttt{christian@cosmos.phy.tufts.edu}}} 

\affiliation{Institute of Cosmology, Department of Physics and Astronomy, Tufts University, Medford, MA 02155, USA}

\pacs{98.80.Cq, 95.36.+x}

\begin{abstract}

A model-independent method to study the possible evolution of dark energy is presented.
Optimal estimates of the dark energy equation of state $w$ are obtained from current 
supernovae data from [A.~G.~Riess {\it et al.}, Astrophys.\ J.\  {\bf 607}, 665 (2004).] 
following a principal components approach.
We assess the impact of varying the number of piecewise constant $w$ estimates 
$N_\mr{p}$ using a model selection method, the Bayesian information criterion, and
compare the most favored models with some parametrizations commonly used in the 
literature.
Although data seem to prefer a cosmological constant, some models are only moderately 
disfavored by our selection criterion: a constant $w$, $w\propto a$, $w\propto z$ and the 
two-parameter models introduced here.
Among these, the models we find by optimization are slightly preferred. 
However, current data do not allow us to draw a conclusion on the possible evolution of dark energy.
Interestingly, the best fits for all varying-$w$ models exhibit a $w<-1$ at low redshifts.
\end{abstract}

\date{\today}

\maketitle

\section{Introduction}
\label{intro}
The discovery of the late-time accelerated expansion of our universe using supernovae (SNe) observations \cite{RP} 
and independent corroboration coming from additional observational tests \cite{sdss}, have often been interpreted 
as indications of the existence of a previously undetected \textit{dark energy}.
Furthermore, this appears to be the dominant constituent of the universe, accounting for $70\%$ of its total 
energy density.
Our understanding of the nature of dark energy is severely limited by the little we know about the dependence on 
the redshift $z$ of its equation of state parameter $w(z)$, defined as the ratio between its pressure and its
energy density.
In practice, observations are analyzed while assuming a particular form of $w(z)$, $w=\mr{const}$ 
being the most common one.
More elaborate parametrizations have been designed to capture the different behaviors of dark energy models,
yet none provides a standard description for $w$.
Among their usual deficiencies are: the lack of flexibility to cope with rapid evolution (as pointed 
out in \cite{BCK}), biasing caused by imposing priors over model parameters, and model-dependent results.
For example \cite{MBMS} warns about the pitfalls of assuming a constant $w$ or $w(z)\geqslant -1$ which,
respectively, may cause an erroneous reconstruction remarkably consistent with observational constraints or
an underestimation of uncertainties.
Studies of possible theoretical motivations for $w(z)<-1$ are available in the literature (see for example 
\cite{MMOT} and references therein).

Principal components (PC) analysis was first used to address the problem of parametrizing $w(z)$ in \cite{HS},
more recent examples of the study of dark energy through PC being \cite{HC, LH, CP,ST}.
In such approach $w(z)$ is described in terms of a basis of orthogonal functions whose form is dictated by 
the constraining capabilities of data.
A shortcoming of employing PC in this context is that even the best determined estimates present broad, 
difficult-to-interpret component functions \cite{T2002}, which obscures the intuitive interpretation of the resulting 
estimates.
On the other hand, the \textit{localized} principal components (LPC) approach taken by Huterer and Cooray
in \cite{HC} benefits from a more straightforward interpretation: the equation of state is now described by 
\textit{uncorrelated} and \textit{localized} piecewise constant estimates at different redshift regions.
Still, the parametrization through LPC is not completely model-independent since the redshift bin arrangement needs to be
determined by hand.
In Ref. \cite{HC} the choice was made on the assumption that better constraining capabilities are attained at lower redshifts.
As will be seen later, we replace this assumption by a detailed inspection of the possible redshift bin configurations.
Namely, we are asking the following question:
Is there an optimal structure of the piecewise constant $w(z)$ that could facilitate its interpretation, with respect to a given 
dataset?
An optimized reconstruction of this sort would provide useful information on the phenomenology of dark energy, together 
with a model independent account of the constraining power of the dataset under consideration.

In this paper we present our efforts to construct such a parametrization:
In Section \ref{method} we show the methodology leading to an optimal LPC description of the equation of state.
First we detail the procedure to obtain a piecewise constant $w$ whose construction is guided by data.
Since we are unable to decide, on theoretical grounds, what number of segments (parameters) to employ in our piecewise 
constant $w(z)$, we recur to a model selection framework to probe the space of possible parametrizations.
This and other statistical tools used here are also described in Section \ref{method}.
Our results and discussion are presented in Section \ref{results}, where the point is made that both the one- and 
two-parameter optimized LPC models are equally suitable descriptions under the model selection criterion and 
dataset considered.
Concluding remarks are presented in section \ref{conc}.
\section{Methodology} \label{method}

\subsection {Optimized piecewise constant \textit{w}(\textit{z})} \label{opcw}

We choose to represent $w(z)$ by a piecewise constant function, assigning a value $w_i$ at each redshift 
interval $(z_{i-1},z_i)$. This is expressed simply as:
\be \label{eq:exp}
w(z) = \sum^{N_\mr{p}}_{i=1} w_i \, b_i(z),
\ee
where a function $b_j(z)$ equals unity within the bin $(z_{j-1},z_j)$ and zero elsewhere. 
The number of segments $N_\mr{p}$ is defined to be the number of model parameters.

A model will also be determined by two nuisance parameters, the matter density $\Omega_m$ and 
the present day normalized Hubble expansion rate $h$.
In addition, to fully determine a model we need to specify the redshift interval boundaries $\{z_\mr{div}\}$.
The freedom in choosing the latter generates a whole family of parametrizations for fixed $N_\mr{p}$.
For example, when restricted to $N_\mr{p} = 2$ we could set the boundaries at $\{0, 0.9, 1.8\}$ 
which would imply different data-fitting capabilities than the choice $\{0, 0.3, 1.8\}$.
From now on, $z_\mr{div}$ refer exclusively to intermediate divisions since our analyses are all for the $(0, 1.8)$ 
range.

In our examples we perform a thorough inspection of the fitting properties of all possible $\{z_\mr{div}\}$ 
configurations.
The optimal model for each $N_\mr{p}$ is identified according to the following outline:
We start by finding the most likely $w_i$ values, with respect to some dataset, for all models in a family.
We then identify the optimal case, the one with highest likelihood (defined below),
and refer to it as the \textit{optimized} model for the $N_\mr{p}$ family, or simply, the $N_\mr{p}$-parameter 
model.
This optimization procedure fully exploits the descriptive power of piecewise constant parametrizations,
since it ensures a data-driven reconstruction of $w$.

The likelihood, $\mathcal{L} = e^{-\chi^2/2}$, follows from the $\chi^2$ statistic, which is in turn given by
 \be \label{chisq}
 \chi^2(\mathbf{p}) = \sum^N_{i=1} \frac{(y^\mr{obs}_j - y^\mr{model}_j(\mathbf{p}) )^2}{\sigma_j^2},
 \ee
where $y^\mr{obs}$ is some observed physical quantity, $\sigma$ its corresponding uncertainty, 
$y^\mr{model}$ the quantity calculated assuming some model given by parameters $\mathbf{p}$
and the sum is over $N$ datapoints.

\subsection {Localized principal components}

As mentioned in Section \ref{intro}, our aim is to describe $w$ in terms of a basis in which parameter errors are uncorrelated
and eigenfunctions are visually easy to interpret.

The standard PC analysis provides, as an alternative to Eq. \ref{eq:exp}, an expansion
\be
w(z) = \sum^{N_\mr{p}}_{i=1} q_i \, e_i(z),
\ee
where the new parameters $\mathbf q$ present uncorrelated errors.
Nevertheless, the window functions $\mathbf e$ ---the eigenfunctions of the decorrelation matrix relating the set of 
original parameters to the uncorrelated ones--- are in general difficult to interpret.
They are often oscillatory and non-zero for the whole redshift range under consideration (\textit{i.e.} not localized).
This motivates the use of the LPC approach, which yields highly localized and mostly positive window 
functions that allow a direct interpretation of the resulting estimates.
As pointed out in \cite{T}, this is a consequence of choosing the square root of the Fisher matrix as the decorrelation 
matrix.

We obtain LPC estimates for our piecewise constant models following the prescription of \cite{HC}, as described in 
the Appendix.

\subsection {Priors}
As mentioned earlier, setting priors on $w(z)$ may decrease our chances of retrieving the correct model.
Therefore, we only use the top-hat prior implicitly defined by the size of the parameter space 
which, when chosen to be large compared to likelihood widths, resembles a `no priors' situation.
Later we will see that ranges with considerable fractions in the $w<-1$ sector are commonplace.

Since our ability to constrain $w(z)$ depends strongly on our knowledge of $\Omega_m$ \cite{LH2003,MBMS},
a tight prior will help us attain optimal model-recognition capabilities \cite{SWB}.
Considering this we marginalize using a somewhat optimistic flat prior $0.20 < \Omega_m < 0.28$, in agreement with 
the 1$\sigma$ constraints of the three-year analysis of WMAP \cite{wmap3}.
Additionally, the Hubble parameter is assigned a flat prior $0.67 < h < 0.79$, compatible with 2$\sigma$ constraints
from WMAP.
A flat universe is assumed throughout.

\subsection{Observational test}

The observational quantity used to compute $\chi^2$ through Equation (\ref{chisq}) is the one usually reported by
experiments: the distance modulus $\mu(z) = m(z) - M_\mr{obs}$ associated to type Ia supernovae.
The apparent magnitude, assumed here to have Gaussian errors, is given by
\be
m(z) = M - 5 \log_{10} h + 5 \log_{10} D_\mr{L} + 42.38.
\ee
while the Hubble-parameter-free luminosity distance is defined as
\be
D_\mr{L} = H_0 \, (1+z) \int^z_0 \frac{du}{H(u)},
\ee
$M$ is the absolute magnitude of supernovae, $H(z)$ the Hubble expansion rate and $H_0$ its present day value.
The perfect degeneracy between $M$ and $h \equiv 10^{-2} H_0\,\mr{km}^{-1}\,\mr{s}\,\mr{Mpc}$ is what allows us to 
reduce the set of nuisance parameters to $\{\Omega_m , h\}$.
An enlightening discussion on the subtleties involved in pairing the theoretical and observed $m(z)$ can be found 
in Ref. \cite{CP2}.

For this analysis we use 156 supernovae from the \textit{Gold} dataset of Riess \etal \cite{gold}, which includes 9 SNe at $z > 1$.
Our $z_\mr{div}$ arrangements are then restricted to contain at least 10 SNe in their last bin, implying 
$z_\mr{div}\lesssim 1$.

\subsection {Model comparison}

As noted in \cite{L}, a model selection approach is essential when comparing different theoretical models in the light 
of observations; in contrast with just exhibiting estimated values and corresponding confidence levels of the model 
parameters associated to the various models.
We take this approach to assess the data-fitting merits of different $N_\mr{p}$ models independently of the number of
model parameters.

The problem of choosing the most probable model, given certain dataset, finds a simple solution with the Bayesian 
information criterion (BIC, \cite{S}):
\be
\mr{BIC}= -2 \ln \mathcal{L}_\mr{max} + N_\mr{p} \ln N_\mr{data},
\ee
where $\mathcal{L}_\mr{max}$ is the maximized likelihood.
Recent examples of the use of the BIC in a cosmological context can be found in \cite{L, BCK, bic} 
among other sources.
This statistical tool was designed as an approximation to the Bayes factor $B$ ---the ratio of posterior likelihoods 
used in Bayesian statistics to compare a pair of competing models--- resulting in the approximate relation 
$\mr{BIC} \sim -2\ln B$.
An exact solution requires calculating the Bayesian evidence \cite{M}, a more sophisticated task involving integration 
of the likelihood-prior product over the parameter space for each model (for an implementation of Bayesian evidence to 
determine the order of a polynomial $w(z)$ see \cite{SWB}).

When priors over model parameters are unknown, as is our case, the BIC is a convenient alternative to the Bayesian 
evidence since it is, by design, independent of parameter space size.
Models may be ranked according to the difference $\Delta$BIC with respect to a base model, chosen here to be the 
`zero-parameters' $\Lambda$CDM.
As listed in Ref. \cite{KR}, a $\Delta$BIC value between $2-5$ ($5-10$) translates into positive (strong) evidence in favor 
of the model with smaller BIC value, while $\Delta$BIC $> 10$ implies that the evidence is decisive. 
A $\Delta$BIC $< 2$ means neither model is preferred by the criterion.

\section{Results and Discussion}\label{results}

First let us define $\Lambda$CDM as the base model for model selection.
The next simplest possibility, constant $w$, is equivalent to the $N_\mr{p}=1$ model, for which we find
an estimate $w_1 = -0.90 \pm 0.12$ (68$\%$ confidence level).

\begin{figure}[htb]
\centering
\includegraphics[width=8cm]{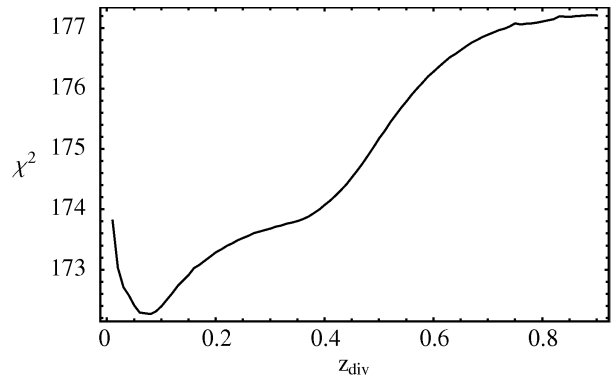}
\caption{Dependence of $\chi^2$ on the bin boundary $z_\mr{div}$ that characterizes LPC models with two segments.
}
\label{fig:Chi2_2p}
\end{figure}

We now illustrate the exploration that identifies the optimal bin arrangement for fixed $N_\mr{p}$.
Fig. \ref{fig:Chi2_2p} shows the dependence of $\chi^2$ on the bin boundary position $z_\mr{div}$ for $N_\mr{p}=2$ 
models.
We identify the model with $z_\mr{div}=0.08$ as the one with a smaller $\chi^2$ and label it as the `optimized model'.
Large-$\chi^2$ (worse fit) models are limiting cases resembling the constant $w$ model, their $\chi^2$ and preferred 
$w_1$ values tend to the ones of the single-parameter case (later shown in Table \ref{tab:models}).
This is not surprising if we bear in mind that $95\%$ of SNe belong to the first redshift bin when $z_\mr{div} \sim 1$.

\begin{table*}[thb]
\caption{
Parameter estimates together with $\chi^2$ and BIC values for various models.
The difference $\Delta$BIC is calculated with respect to the base model $\Lambda$CDM.
Redshift arrangements are shown for our optimized models.}
\begin{tabular*}{0.9\textwidth}{@{\extracolsep{\fill}} lc cc rrrrr ccr}
\hline
\hline
Model& Optimized $\{z_\mr{div}\}$ &$w_0$&$w_{a,z}$&$q_1$&$q_2$&$q_3$ &$q_4$&$q_5$&$\chi^2_\mr{min}$ &BIC &$\Delta$BIC\\
\hline
 $\Lambda$CDM  &                                      &$\equiv$-1&&          &          &             &             &           & 178.6 &178.6 &0.0 \\
 $N_\mr{p}=1$ &                                                &&&-0.90&            &             &            &           & 177.9 &182.9  & 4.4  \\
 $N_\mr{p}=2$ &  $\{0.08\}$                             &&&-2.05 &-0.60  &            &             &           & 172.7  &182.8 & 4.3   \\
 $N_\mr{p}=3$ &  $\{0.43, 0.60\}$                   &&&-1.08 &2.01   &-1.90    &             &           & 171.2 &186.4 & 7.8   \\
 $N_\mr{p}=4$ &  $\{0.05, 0.55, 0.70\}$         &&&-3.56 &-0.63  &7.45     &-25.01  &           & 170.5 &190.7 & 12.1 \\
 $N_\mr{p}=5$ &  $\{0.05,0.50,0.60,0.70\}$  &&&-3.62  &-0.67 &9.80     &-3.94    &-40.16& 169.6 &194.8 & 16.3  \\
 eq. err.        & $\{ 0.18 \}$                           &&&-1.28 &-0.46  &             &             &           & 173.4 &183.5 & 4.9  \\
 $w\propto a $  &                                          &-1.32&2.61&&    &             &             &           & 173.8 &183.9 & 5.3    \\
 $w\propto z$   &                                          &-1.27&1.63&&    &             &             &           & 174.3 &184.4 & 5.8   \\
  
  \hline
  \hline
\end{tabular*}
\label{tab:models}
\end{table*}
\begin{figure}[!ht]
\centering
\begin{tabular}{c}
\includegraphics[width=7cm]{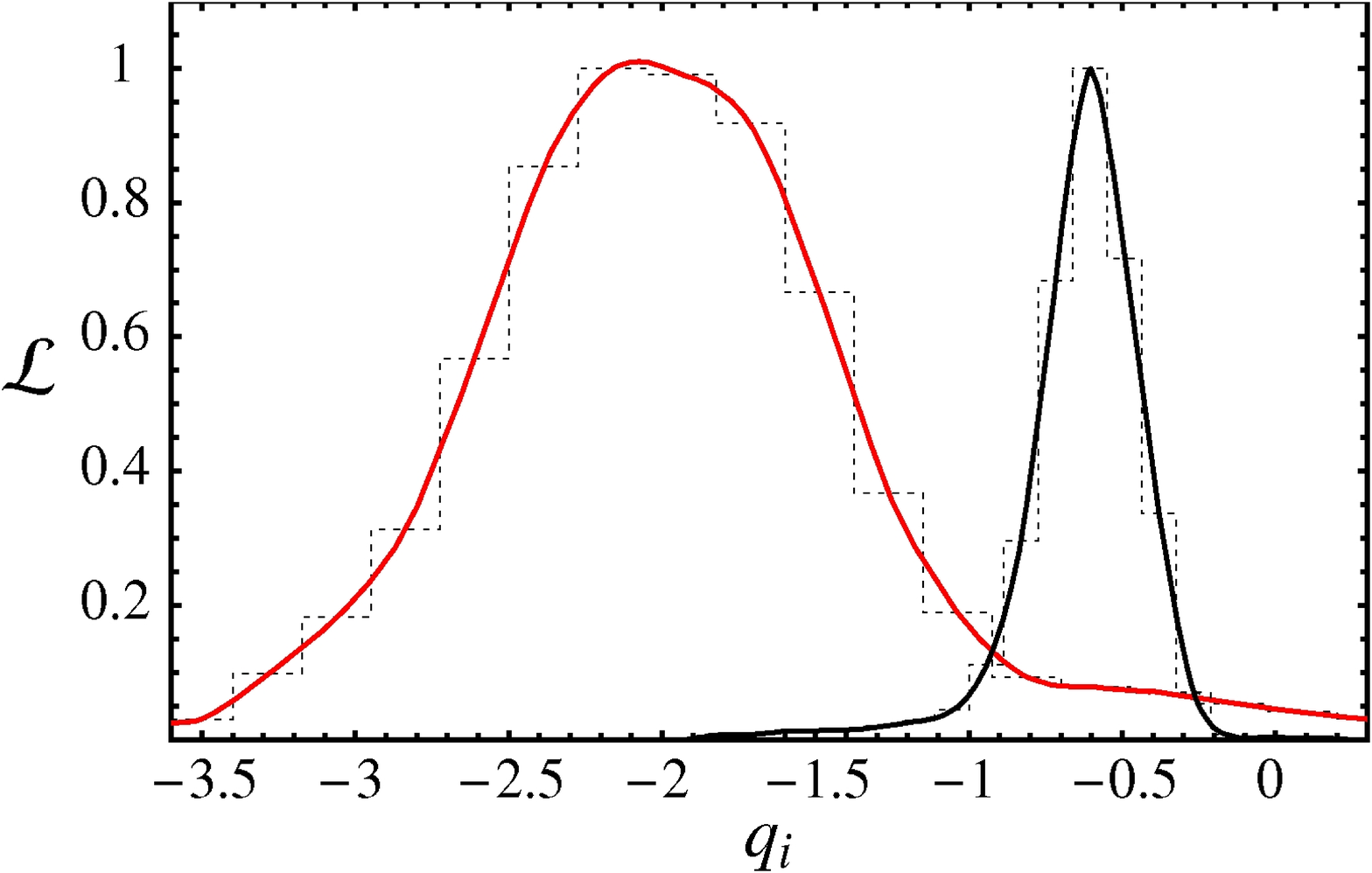}\\
\includegraphics[width=7cm]{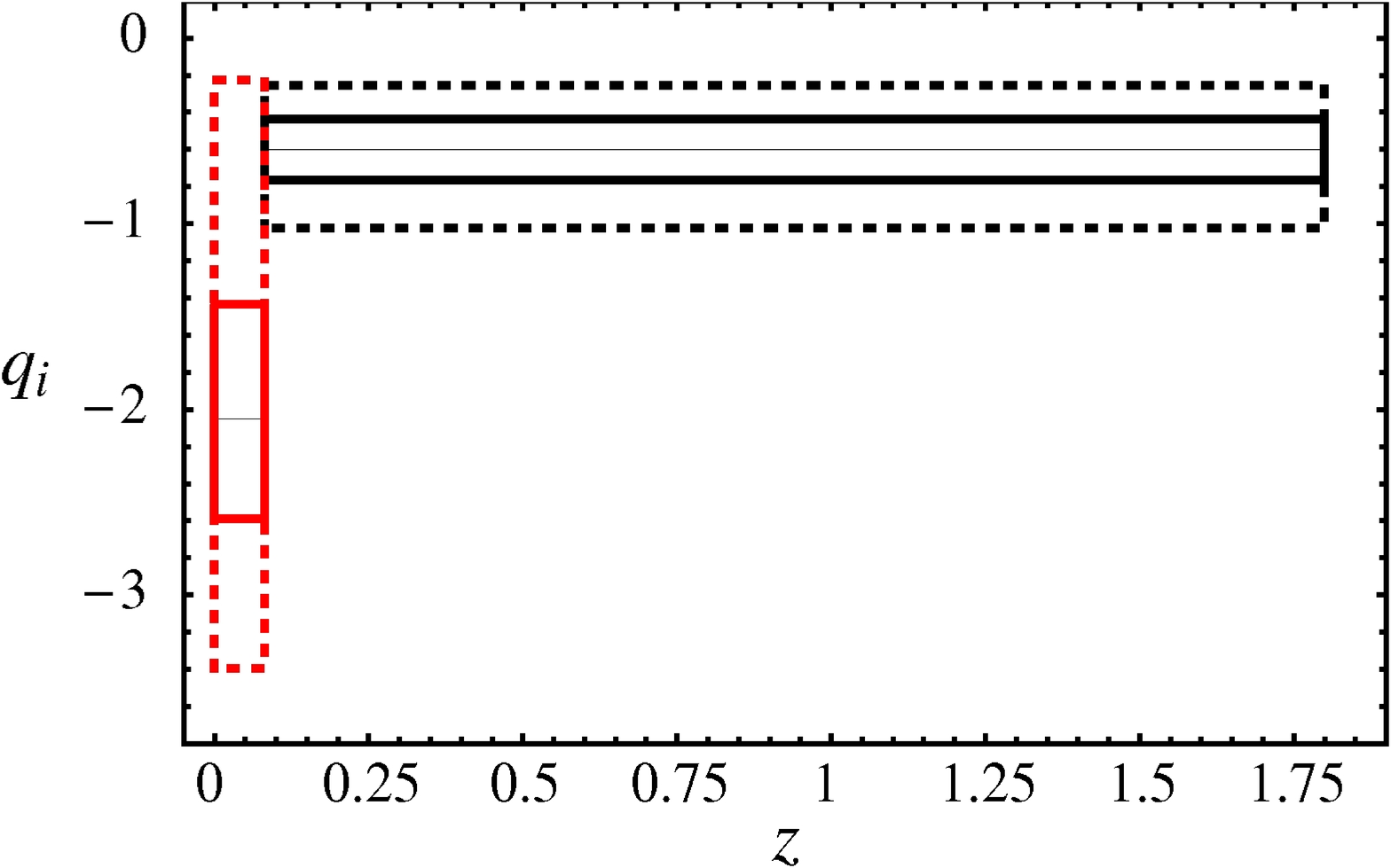}\\
\includegraphics[width=7cm]{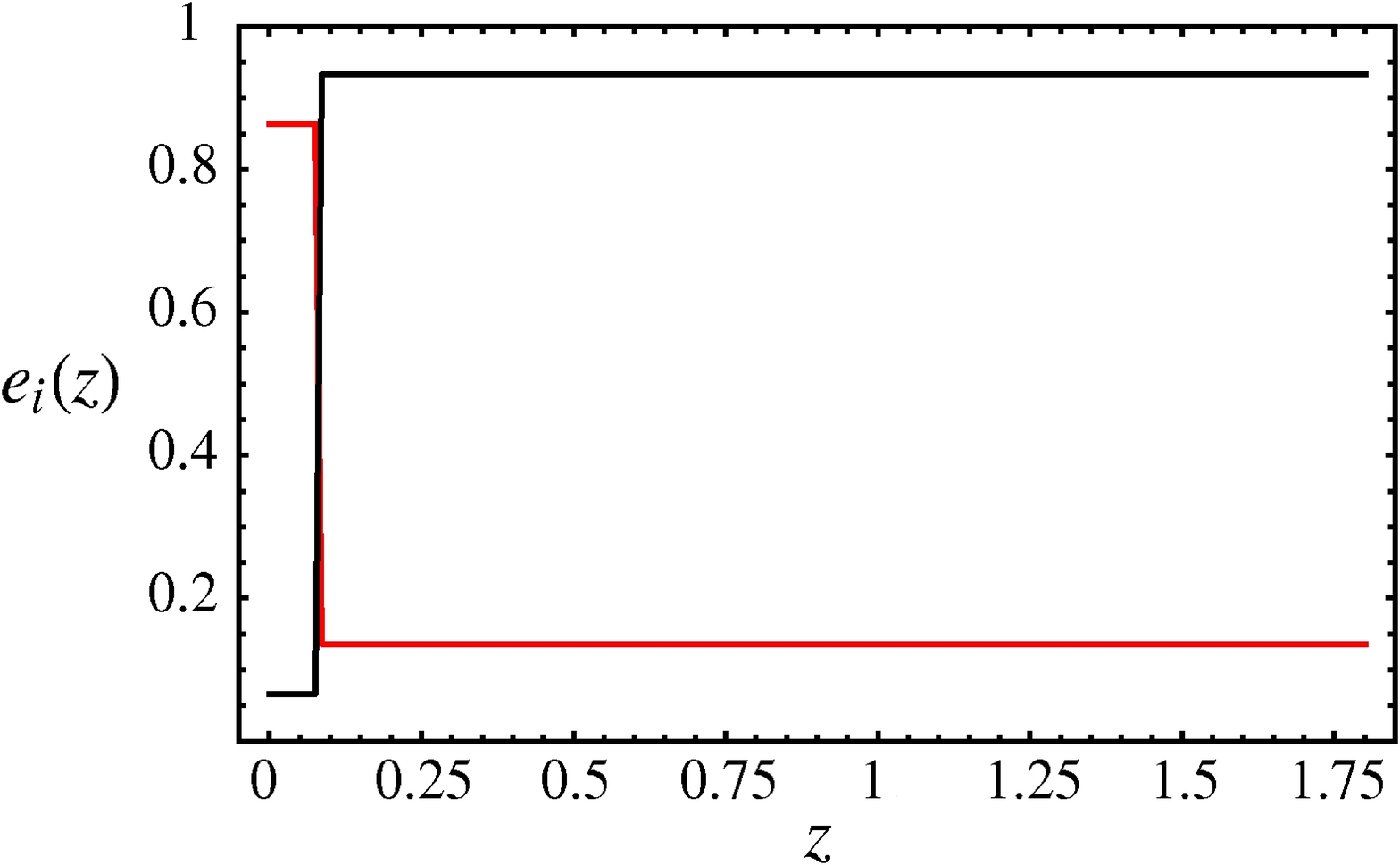}\\
\end{tabular}
\caption{ (color online)
The optimized $N_\mr{p}=2$ model.
Top: Likelihoods for the optimized 2-parameter model.
Note that the low-redshift estimate (red/gray line) is broader than the medium- and large-redshift estimate (black line).
Color conventions are the same for the three plots.
Middle: Uncorrelated principal component estimates of $w(z)$ (thin, horizontal lines) with corresponding 1-$\sigma$ (solid) 
and 2-$\sigma$ (dotted) rectangles showing parameter constraints (horizontal edges) and redshift span (vertical edges).
Bottom: Window functions relating original piecewise coefficients to LPC ones.}
\label{fig:W3_2p}
\end{figure}
In Table \ref{tab:models} we present the models obtained by means of optimization for $N_\mr{p}=1-5$, together with 
the base model $\Lambda$CDM and two other parametrizations commonly used in the literature:
$w(z) = w_0 + w_z z$ \cite{wz} and $w(a) = w_0 + w_a (1-a)$ \cite{wa}.
We also include the model selection quantity $\Delta$BIC, all statistics are obtained using the \textit{Gold} dataset.

Model comparison in light of the BIC criterion shows that $\Lambda$CDM is the model preferred by data.
A collection of models that exhibit low $\chi^2$ but are moderately disfavored are $N_\mr{p}=1$, $N_\mr{p}=2$,
$w \propto z$, $w \propto a$ and the \textit{equal errors} model explained below.
The $N_\mr{p}=3$ model turns out to be strongly disfavored while $N_\mr{p}=4$ and $N_\mr{p}=5$ are 
decidedly disfavored, all this in terms of the rules proposed in \cite{KR}.

It is worth mentioning that all time-varying $w$ models, favored by the BIC or not, suggest a principal component value $q_1 < -1$ at low redshifts.
The table also suggests that a larger number of LPC estimates implies more extreme values for $q_i$.
This reminds of certain parametrizations that, when extended to second order expansions, acquire huge values 
in their expansion coefficients, as studied in \cite{BCK}.
We won't be concerned by the huge values found, since the models in question are already rejected by the BIC criterion. 
Still, it is somewhat disconcerting that the further redshift divisions are added, estimates are more and more inconsistent with
$q_1\sim-1$.

A more detailed analysis of the optimized $N_\mr{p}=2$ case shows that $q_1$ is loosely constrained with respect to 
$q_2$.
This is illustrated in Figure \ref{fig:W3_2p}, where likelihoods (top plot) and constraints on model parameters 
$q_i$ (middle plot) are shown.
The bottom plot of the figure shows the highly localized, mostly positive window functions resulting from the LPC 
approach.

\begin{figure}[hbt]
\centering
\includegraphics[width=7.5cm]{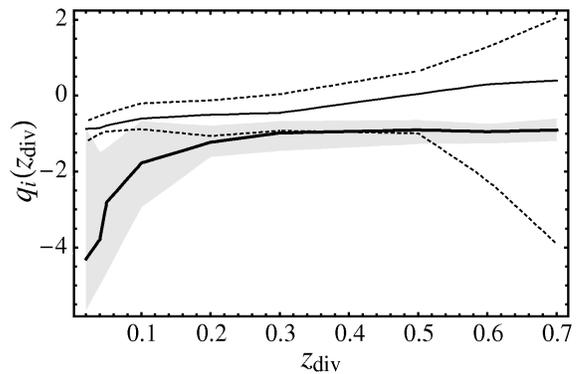}
\caption{Parameter estimates $q_1$ (thick line) and $q_2$ (thin line) for varying $z_\mr{div}$ and their corresponding
2$\sigma$ confidence regions.
}
\label{fig:zvse}
\end{figure}

\begin{figure}[!ht]
\vspace{.8cm}
\centering
\begin{tabular}{c}
\includegraphics[width=7cm]{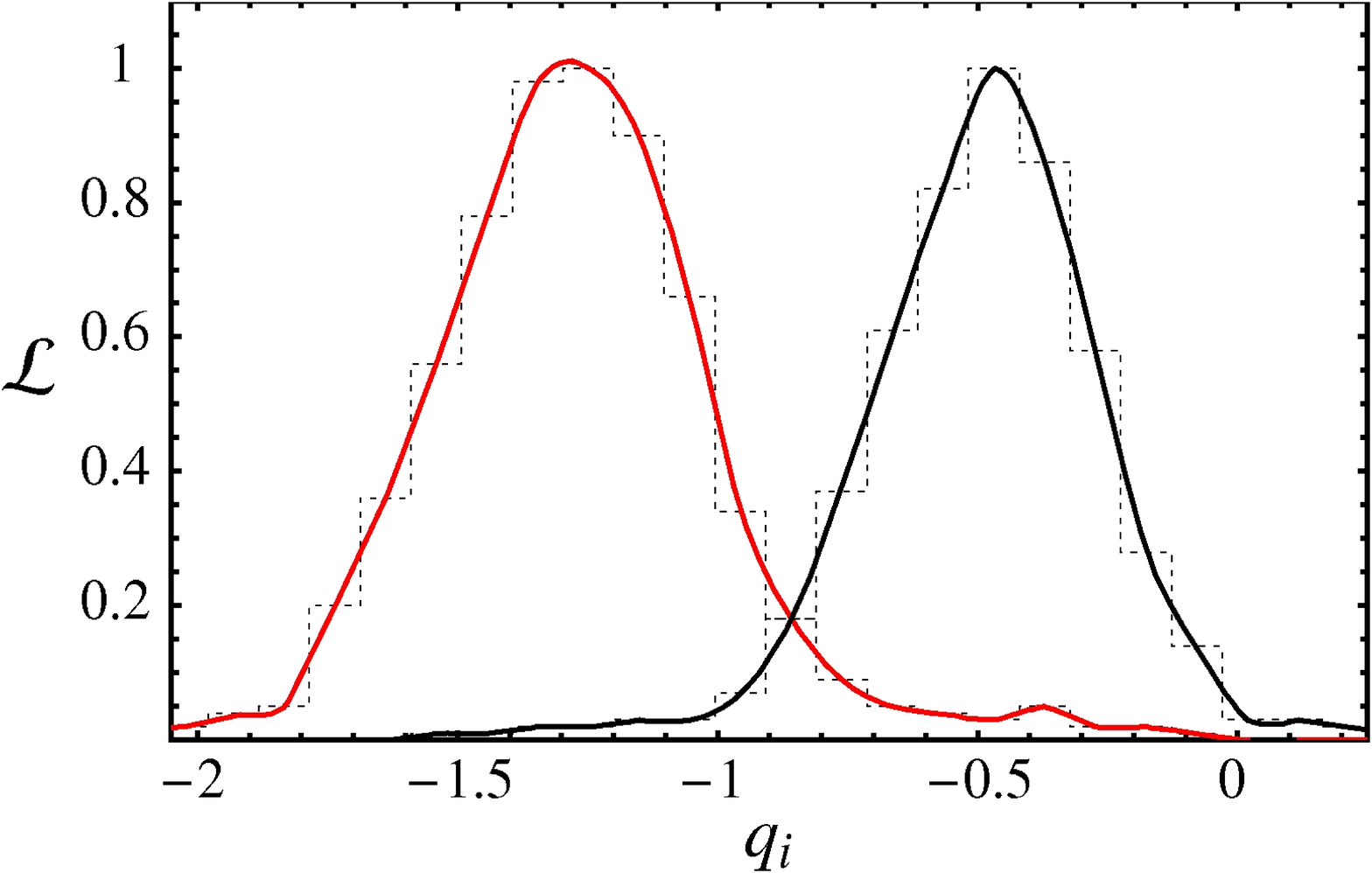}\\
\includegraphics[width=7cm]{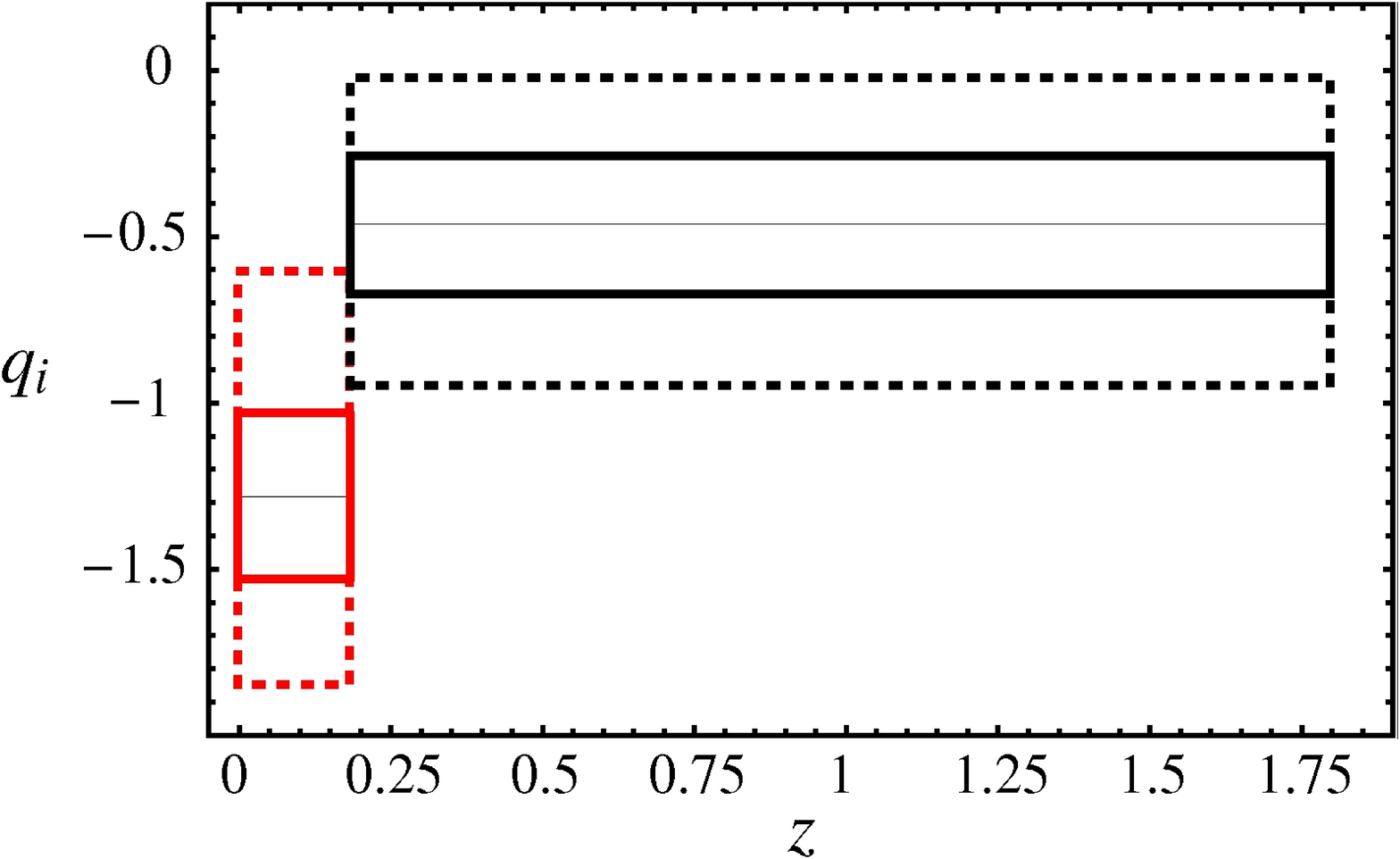}\\
\includegraphics[width=7cm]{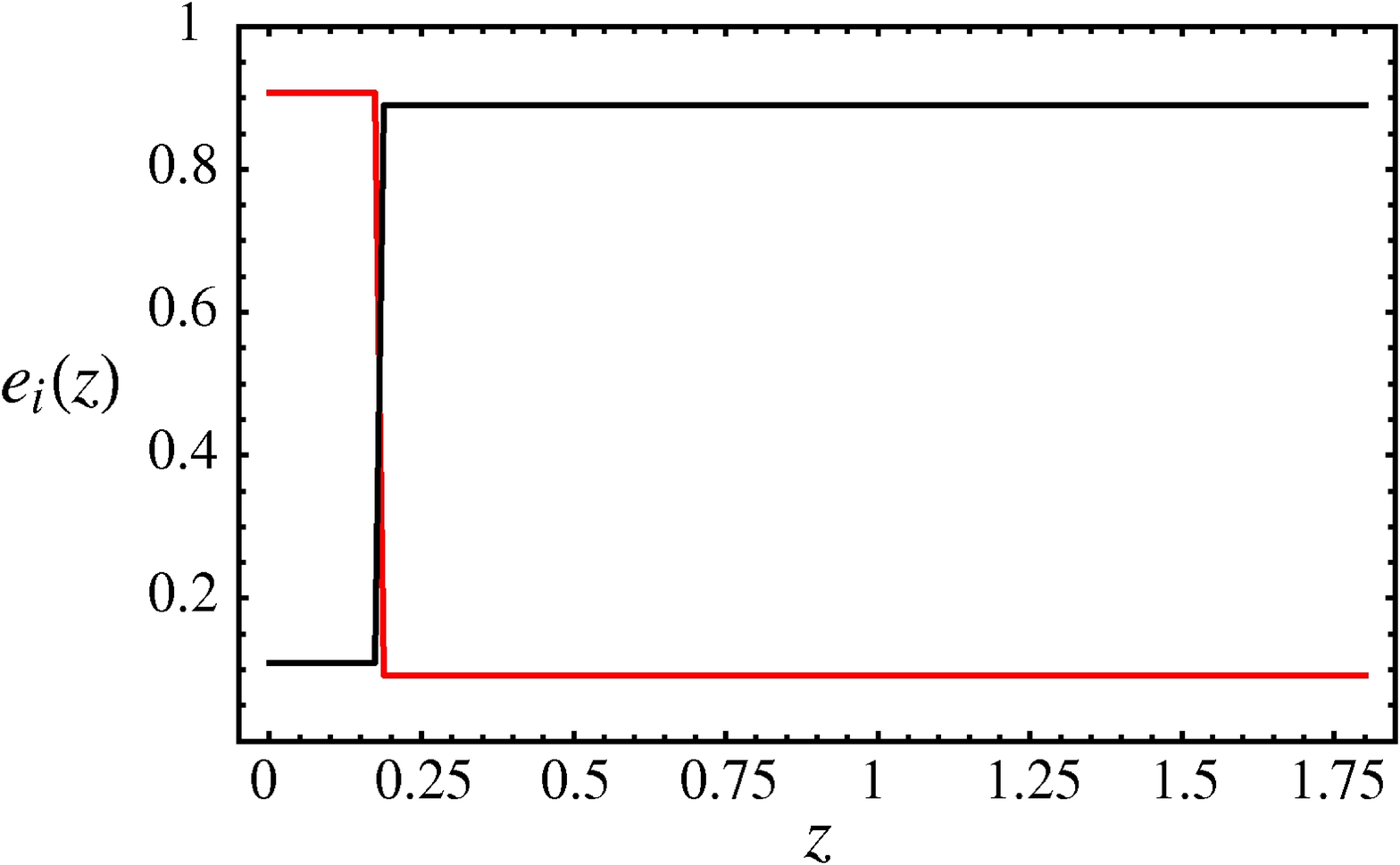}\\
\end{tabular}
\caption{ (color online)
The \textit{equal errors} model, a 2-parameter model similar to the optimized one but with comparable errors on 
both $q$-estimates.
Likelihoods (top),  $q$-estimates describing $w(z)$ (middle) and window functions (bottom) are displayed following 
the color coding of Figure \ref{fig:W3_2p}.}
\label{fig:e_2p}
\end{figure}

An alternative $N_\mr{p}=2$ model is obtained by a different optimization criterion which, instead of accepting the model 
with minimum $\chi^2$, requires errors in parameter estimates to be of comparable size.
This case exemplifies how different optimization criteria may result in different $w$-estimates.
In figure \ref{fig:zvse} we sketch the interplay between $q_1$ and $q_2$ estimates when $z_\mr{div}$ is varied. 
A detailed analysis points to the $z_\mr{div}=0.18$ model as the one with matching errors, which is
labeled the \textit{equal errors} model. 
As seen in Table \ref{tab:models}, this model has practically the same BIC value as the $N_\mr{p}=1$ and $N_\mr{p}=2$ 
models.
Its corresponding likelihood, parameter estimates and window functions are plotted in Figure \ref{fig:e_2p}.

It is interesting that two of the models with good fits to the data, the optimized $N_\mr{p}=2$ and 
\textit{equal errors} ones, are reminiscent of some more complex models ---for example the ones in 
\cite{BCK,HC,mod}--- which suggest $w< -1$ at low redshift and $w\sim-0.5$ elsewere.
However, our models rely on two parameters only, which makes them more attractive in terms of model selection, 
making a better case for the rapidly varying $w$ possibility.


\section{Conclusions} \label{conc}
This paper introduces a  new framework to study the 
model-independent implications of data regarding the possibility
of dark energy evolution.
Extending the approach proposed in \cite{HC}, we find 
the optimal set of eigenfunctions that describe $w$.
We focus on a currently available SNe dataset \cite{gold},
and follow a model selection scheme, in terms of the Bayesian 
information criterion, which points to $\Lambda$CDM as the 
preferred model.
A class of models is found to provide better fits to data but is 
moderately disfavored by the BIC. 
These include our two-parameter models: the one that 
minimizes $\chi^2$ and the one with equal errors on both 
estimates, as well as the constant $w$ parametrization.
Also in this category, with slightly worse BIC values, are the 
`linear in redshift' and `linear in scale factor' parametrizations.
Optimized models with more than two parameters are strongly 
disfavored.

In the light of the data considered, mutually exclusive 
phenomenologies are found to be equally favored by the BIC, 
for example, the constant $w$ and optimized $N_\mr{p}=2$ 
models which are, respectively, consistent and marginally 
consistent with $\Lambda$CDM at a 95$\%$ confidence 
level.
This reproduces a problem commonly found in the literature
when considering flexible forms of $w(z)$: the constant 
$w$ model  leads to $w\sim-1$ at all redshifts, but if $w$ is 
allowed to evolve we find that $w<-1$ at low redshift and 
$w>-1$ elsewhere.
Our approach may prove useful in reconstructing a true 
model with a rapidly varying $w$, however, current data give 
no information on whether this option is preferred over a slowly 
varying or constant $w$, even when the rapidly varying $w$ is 
described with a simple model such as our two-parameter ones.

Interpretation of the favored models should be made while considering 
that resulting estimates are ``integrated properties of w(z)'' \cite{SPB}, 
and so the underlaying physical model may differ substantially from 
the reconstructed one.
In particular, the $z_\mr{div}$ found in our optimized models is not 
necessarily a proof of change in dark energy dynamics at exactly that 
time, it could be seen as a feature of the reconstruction employed.

Further research, adding other observational tests to the analysis, may 
improve constraints on the equation of state.
Of special interest are weak lensing measurements \cite{wl}, or the observed 
Integrated Sachs-Wolfe effect \cite{SW}, which for certain cases provides a 
probe of the total change in $w$ \cite{PCSCN}.
This is postponed for a future investigation.

\section*{Acknowledgments}
It is a pleasure to thank Dragan Huterer for very useful correspondence, Jon Urrestilla and Martin Sahl\'en for their helpful comments and specially Levon Pogosian for very insightful discussions.
I also thank the referee for pointing out reference \cite{wz}.

\appendix*
\section{}
In this appendix we show, following \cite{HC}, the procedure to decompose $w$ in terms of localized principal 
components.

We first calculate the Fisher matrix $\mathbf{F}$ \cite{TTH} (marginalized over nuisance parameters) for some $w$ model 
and with respect to a given dataset.
We then obtain its square root
\be
\mathbf{F}^{1/2}= \mathbf{O}^\mr{T} \sqrt{\mathbf{\Lambda}}\mathbf{O},
\ee
where the diagonalized matrix of eigenvalues $\mathbf{\Lambda}$ and the orthogonal matrix 
of eigenvectors $\mathbf{O}$ satisfy
\be
\mathbf{F}= \mathbf{O}^\mr{T} \mathbf{\Lambda} \mathbf{O}.
\ee
It is straightforward to check that $\mathbf{F}^{1/2}$ is indeed the `square root matrix' of $\mathbf{F}$ if we 
consider the product $\mathbf{F}^{1/2}\mathbf{F}^{1/2}$.

LPC estimates $\mathbf{q}$ are given in terms of the original parameters $\mathbf{p}$ by 
$\mathbf{q}=\mathbf{W} \mathbf{p}$, where $\mathbf{W}$ is a normalized version of $\mathbf{F}^{1/2}$ 
for which each row adds up to 1 (Roman indices run from 1 to $N_\mr{p}$ and summations are explicitely 
shown):
\be
W_{ij} = \frac{F^{1/2}_{ij} }{ \sum_k \limits F^{1/2}_{ik}}.
\ee
The rows of $W$ then define the window functions $\mathbf{e}$ which provide the desired orthogonal basis.
Furthermore, when considering the covariance of the original parameters 
$\langle \Delta p_i \Delta p_j \rangle = F^{-1}_{ij}$ and the diagonality of $W$, the new errors $\Delta q_i$ 
are found to be uncorrelated, their covariance being
\be
\langle \Delta q_i \Delta q_j \rangle = \frac{\delta_{ij}}{\sum_{k} \limits F^{1/2}_{ik}  \sum_{l} \limits F^{1/2}_{jl}}.
\ee
Our expansion in terms of the window functions is then written as:
\be
w(z) = \sum_i q_i \, e_i(z),
\ee
where instead of ordering the principal components, as is usual, according to the magnitude of individual 
uncertainties, we respect the original indexing to maintain the transparency of the interpretation.

\end{document}